\documentclass {elsart}
\usepackage{graphicx}

\usepackage{graphics}
\usepackage{graphicx}
\usepackage{epsfig}

\usepackage{amssymb}
\usepackage{amsmath}

\begin{document}

\begin{frontmatter}



\title{Scaled Bregman divergences in a Tsallis scenario}


\author[SRC]{R. C. Venkatesan\corauthref{cor}}
\corauth[cor]{Corresponding author.}
\ead{ravi@systemsresearchcorp.com}
\author[UNLP]{A. Plastino}
\ead{plastino@venus.fisica.unlp.edu.ar}

\address[SRC]{Systems Research Corporation,
Aundh, Pune 411007, India}
\address[UNLP]{IFLP, National University La Plata \&
National Research Council (CONICET)\\ C. C., 727 1900, La Plata,
Argentina}

\begin{abstract}
There exist two different versions  of the Kullback-Leibler
divergence (K-Ld) in Tsallis statistics, namely the usual
generalized K-Ld and the generalized Bregman K-Ld. Problems have
been encountered in trying to reconcile them. A condition for
consistency between these two generalized K-Ld-forms by recourse to
the additive duality of Tsallis statistics is derived. It is also
shown that the usual generalized K-Ld subjected to this additive
duality, known as the dual generalized K-Ld, is a scaled Bregman
divergence. This leads to an interesting conclusion: the dual
generalized mutual information is a scaled Bregman information. The
utility and implications of these results are discussed.

\end{abstract}

\end{frontmatter}

\section{Introduction}
The generalized statistics of Tsallis' has recently been the focus
of much attention in statistical physics, complex systems, and
allied disciplines (in this paper the terms generalized statistics,
nonadditive statistics, and nonextensive statistics are indistinctly
used)[1]. It is well-known that nonadditive statistics generalizes
the extensive Boltzmann-Gibbs-Shannon (B-G-S) statistics. Its scope
has lately been extended to studies of lossy data compression in
communication theory [2] and machine learning [3,4]. In this paper,
attention is focussed upon the Tsallls-generalization of the concept
of relative entropy, also known as  Kullback-Leibler divergence
(K-Ld), that constitutes a fundamental distance-measure in
information theory [5]. The generalized K-Ld [6] encountered in
deformed statistics has been described by Naudts [7] both as a
special form of f-divergences [8, 9], and also in terms of
Bregman-divergences [10]. Bregman divergences are, in turn,
information geometric tools that have lately acquired great
significance in a variety of disciplines ranging from information
retrieval [11] and lossy data compression-machine
learning [12] to statistical  physics [13]. \\

The generalized K-Ld is defined as [7]
\begin{equation} \label{one}
D_\phi  \left( {\left. p \right\|r} \right) =  - \sum\limits_i {p_i
} \omega _\phi  \left( {\frac{{r_i }}{{p_i }}} \right) =
\frac{1}{\kappa }\sum\limits_i {p_i } \left[ {\left( {\frac{{p_i
}}{{r_i }}} \right)^\kappa   - 1} \right],
\end{equation}
where $p$ is an arbitrary distribution, $r$ is the reference
distribution, and $\kappa$ is some nonadditivity parameter
satisfying $ - 1 \le \kappa  \le 1;\kappa  \ne 0 $. Here (1) employs
the definition of the so-called \textit{deduced logarithm }[7]
\begin{equation}
\omega _\phi \left( x \right) = \frac{1}{\kappa }\left( {1 - x^{ -
\kappa } } \right) \ .
\end{equation}

An alternate form of the generalized K-Ld derived from the theory of
Bregman divergences [7] is shown to be
\begin{equation} \label{dos}
D_{\phi}^B  \left( {\left. p \right\|r} \right) = S_\phi  \left( r
\right) - S_\phi  \left( p \right) - \sum\limits_i {\left( {p_i  -
r_i } \right)\ln _\phi  } \left( {r_i } \right),
\end{equation}
where the generalized entropy and the deformed logarithm are defined
as
\begin{equation} S_\phi \left( z \right) = \sum\limits_i {z_i \omega
_\phi \left( {\frac{1}{{z_i }}} \right)},
\end{equation}
and
\begin{equation}
\ln _\phi \left( x \right) = \left( {1 + \kappa ^{ - 1} }
\right)\left( {x^\kappa   - 1} \right),
\end{equation}
respectively.

\subsection{Problems reconciling the Tsallis versions of the Kullback-Leibler divergence}
Specializing the above concepts to the Tsallis scenario by setting
$\kappa=q-1$, Eqs. (1) and (3) yield the usual doubly convex
generalized K-Ld [6]

\begin{equation}
D_{K-L}^q  \left( {\left. p \right\|r} \right) = \frac{1}{q-1
}\sum\limits_i {p_i } \left[ {\left( {\frac{{p_i }}{{r_i }}}
\right)^{q-1}   - 1} \right],
\end{equation}
and the generalized Bregman K-Ld
\begin{equation}
D_{q }^B \left[ {\left. p \right\|r} \right] = \frac{1}{{\left( {q -
1} \right)}}\sum\limits_i {p_i } \left( {p_i^{q-1 }  - r_i^{q-1 } }
\right) - \sum\limits_i {\left( {p_i  - r_i } \right)r_i^{q-1  } },
\end{equation}
respectively.

  While
the form of the generalized Bregman K-Ld (BK-Ld) is more appealing
than (6) from an information geometric viewpoint, it does contain
certain inherent drawbacks.

A study by Abe and Bagci [13] has demonstrated that
 the generalized K-Ld defined by (6) is  jointly convex in
terms of both $p_i$ and $r_i$ while  the form defined by (7) is
convex {\it only} in terms of $p_i$. A further distinction between
the two forms of the generalized K-Ld concerns the property of
composability. While the form defined by (6) is composable, the form
defined by (7) does not exhibit this property.  The fact that the
two generalized K-Ld versions have no apparent relation to each
other should be  a cause of concern for practitioners of
nonextensive statistical physics.

 A second issue to address concerns the manner in which mean values are computed.
 Nonextensive statistics has employed a number of forms in which
expectations may be defined. Prominent among these are the linear
constraints originally employed by Tsallis [1] (also known as normal
averages) of the form: $ \left\langle A \right\rangle =
\sum\limits_i {p_i } A_i $, the Curado-Tsallis (C-T) constraints
[14] of the form:  $ \left\langle A \right\rangle _q  =
\sum\limits_i {p_i^q } A_i  \ $, and the normalized
Tsallis-Mendes-Plastino (TMP) constraints [15] (also known as
$q$-averages) of the form:  $ \left\langle {\left\langle A
\right\rangle } \right\rangle _q  = \sum\limits_i {\frac{{p_i^q
}}{{\sum\limits_i {p_i^q } }}A_i } \ $. A fourth constraining
procedure is the optimal Lagrange multiplier (OLM) approach [16]. Of
these four methods to describe expectations, the most commonly
employed by Tsallis-practitioners is the TMP-one.

Recent works by Abe [17, 18] suggest that in generalized
statistics expectations defined in terms of normal averages, in
contrast to those defined by $q$-averages, are consistent with the
generalized H-theorem and the generalized \textit{Stosszahlansatz}
(molecular chaos hypothesis). The correctness of normal average
expectations vis-\'{a}-vis $q$-average (or TMP) ones has also been
investigated by Hasegawa [19, 20].  Understandably,  a
re-formulation of the variational perturbation approximations in
nonextensive statistical physics followed [21], via an application
of $q$-deformed calculus [22].

Further concern is originated by a consistency issue. This stems
from the fact that the form of the generalized K-Ld defined by (6)
is consistent with expectations and constraints defined by
$q$-averages while, on the other hand, the generalized Bregman K-Ld
defined by (7) is consistent with expectations defined by normal
averages [13].

\subsection{Additive duality}

The additive duality is a fundamental property in generalized
statistics.  One implication of the additive duality is that it
permits a deformed logarithm defined by a given nonadditivity
parameter (say, $q$) to be inferred from its \textit{dual
deformed} logarithm [1, 2, 23] parameterized by: $q^*=2-q$.

Our leitmotif for invoking the additive duality stems from the form
of the BK-Ld (7).  Setting $\kappa=q-1$ in (2) and (5) yields a
Tsallis entropy of the  form: $S_{q}(z)=-z\ln_q(z)$, which is the
Tsallis entropy defined in Section 2.1 of this paper subjected to
the re-parameterization $q\rightarrow 2-q$.\footnote {Here "$
\rightarrow $" denotes a re-parameterization of the nonadditivity
parameter, and is not a limit.}  Thus, in the Tsallis scenario, (5)
is actually the dual Tsallis entropy defined in (15) with the
additive duality ($q\rightarrow 2-q$) \textit{implicitly} accounted
for. Given these facts, from the definition of Bregman divergences
provided by Definition 1 in Section 2.3 below, the form of the BK-Ld
(7) can only be obtained by specifying the complex generating
function as: $\phi(z)=z\ln_qz$, followed by the re-parameterization
$q\rightarrow 2-q$.  More specifically, the BK-Ld (7) can only be
derived from first principles using (5) defined in the Tsallis
scenario by recourse to the additive duality. Hence, the necessity
for invoking the additive duality in this paper, where the
re-parameterization is \textit{explicitly} accounted for by
defining: $q^*=2-q$.

By definition (see Section 2.1 below for details), the generalized
K-Ld subjected to the additive duality is referred to as the dual
generalized K-Ld having the form
\begin{equation}
D_{K - L}^{q^*} \left[ {\left. p \right\|r} \right] =  \sum\limits_i
{p_i \ln_{q^*} \left( {\frac{{p_i }}{{r_i }}} \right)}  =
\frac{1}{{\left( {1 - q^ *  } \right)}}\sum\limits_i {\left( {
p_i^{2 - q^ *  } r_i^{q^ *   - 1} -1} \right)}.
\end{equation}
However, employing the definitions of Bregman divergences presented
in  Section 2.3 below, the BK-Ld is of the form
\begin{equation}
D_{q^ *  }^B \left[ {\left. p \right\|r} \right] = \frac{1}{{\left(
{1 - q^ *  } \right)}}\sum\limits_i {p_i } \left( {p_i^{1 - q^ *  }
- r_i^{1 - q^ *  } } \right) - \sum\limits_i {\left( {p_i  - r_i }
\right)r_i^{1 - q^ *  } },
\end{equation}
for the convex generating function: $\phi(z)=z\ln_{q^*}z$.
\subsection{Goal of this paper}

\textit{Scaled Bregman divergences}, formally introduced by Stummer
[24] and Stummer and Vajda [25], unify separable Bregman divergences
[10] (defined below in Section 2.3) and f-divergences [8,9]. This
paper uses scaled Bregman divergences as its basis, and accomplishes
the following objectives:

\begin{itemize} \item $(i)$ the generalized K-Ld defined by (6) subjected to the
additive duality (dual generalized K-Ld (8) and (15)) is shown to be
consistent with the canonical probability that maximizes the dual
Tsallis entropy of the form [2, 26]: $ S_{q^
* } = - \sum\limits_i {p_{i} } \ln _{q^
* } p_{i }$ employed in conjunction with expectations defined by
normal averages (Section 3 of this paper), \item $(ii) $ a
correspondence between the dual generalized K-Ld and the
generalized Bregman K-Ld is derived (Section 4 below),
\item $(iii)$ the dual generalized K-Ld is demonstrated to be a
scaled Bregman divergence and that its expectation is a
\textit{scaled Bregman information}, i.e. the expectation of a
scaled Bregman divergence (Section 5 below) for both regimes of the
dual nonadditivity parameter $0<q^*<1$ and $q^*>1$ [27] (Section 5
below). \end{itemize} Section 6 is devoted to discussion and
conclusions.  The primary conclusion of this paper is the necessity
of employing the dual generalized K-Ld when performing a minimum
cross entropy analysis (principle of minimum discrimination
information) of Kullback [28] and Kullback and Khairat [29] using
constraints defined by {\it normal average expectations}.

\section{Theoretical preliminaries}

The essential concepts around which this communication revolves
are reviewed in the  three subsections that follow.

\subsection{Tsallis entropy and the additive duality}
\ By definition, the Tsallis entropy, is defined in terms of
discrete variables as [1]
\begin{equation}
\begin{array}{l}
S_q \left( X \right) = -\frac{{1 - \sum\limits_x {p^q \left( x
\right)} }}{{1 - q}}; \sum\limits_x {p\left( x \right)}  = 1. \\
\end{array}
\end{equation}
The constant $ q $ is referred to as the nonadditive parameter.
Here, (10) implies that extensive B-G-S statistics is recovered as $
q \to 1 $. Taking the limit $ q \to 1 $ in (10) and invoking
l'Hospital's rule, $ S_q \left( X \right) \to S\left( X \right) $,
i.e.,  the Shannon entropy.  Nonextensive statistics is intimately
related to \textit{q-deformed }algebra and calculus (see [22] and
the references within). The \textit{q-deformed} logarithm and
exponential are defined as [22]
\begin{equation}
\begin{array}{l}
\ln _q \left( x \right) = \frac{{x^{1 - q} - 1}}{{1 - q}}, \\
and, \\
\exp _q \left( x \right) = \left\{ \begin{array}{l}
 \left[ {1 + \left( {1 - q} \right)x} \right]^{\frac{1}{{1 - q}}} ;1 + \left( {1 - q} \right)x \ge 0 \\
 0;otherwise, \\
 \end{array} \right.
\end{array}
\end{equation}
respectively. In this respect, an important relation from
\textit{q-deformed} algebra is  [2, 22, 27]
\begin{equation}
\begin{array}{l}
 \ln _q \left( {\frac{x}{y}} \right) = y^{q - 1} \left( {\ln _q x - \ln _q y} \right). \\
 \end{array}
\end{equation}
The Tsallis entropy (10), conditional Tsallis entropy, and, joint
Tsallis entropy may be written as [1]
\begin{equation}
\begin{array}{l}
S_q \left( X \right) =  - \sum\limits_x {p\left( x
\right)} ^q \ln _q p\left( x \right), \\
 S_q \left( {\left. {\tilde X} \right|X} \right) =  - \sum\limits_x {\sum\limits_{\tilde x} {p\left( {x,\tilde x} \right)^q \ln _q p\left( {\left. {\tilde x} \right|x} \right)} } , \\
 S_q \left( {X,\tilde X} \right) =  - \sum\limits_x {\sum\limits_{\tilde x} {p\left( {x,\tilde x} \right)^q \ln _q p\left( {x,\tilde x} \right)} }  \\
=S_q(X)+S_q(\tilde X|X)=S_q(\tilde X)+ S_q(X|\tilde X),
\end{array}
\end{equation}
respectively.

This paper makes prominent use of the \textit{additive duality} in
nonextensive statistics. Setting $ q^*=2-q $, from (11) the
\textit{dual deformed} logarithm and exponential are defined as
\begin{equation}
\begin{array}{l}
 \ln _{q^*}  \left( x \right) =  - \ln _q  \left( {\frac{1}{x}} \right), and, \exp _{q^*}  \left( x \right) = \frac{1}{{\exp _q  \left( { - x} \right)}}. \\
 \end{array}
\end{equation}

The dual Tsallis entropy, the dual conditional Tsallis entropy, the
dual joint Tsallis entropy , and, the dual generalized K-Ld may thus
be written as
\begin{equation}
\begin{array}{l}
S_{q^*} \left( X \right) =  - \sum\limits_x {p\left( x
\right)}  \ln _{q^*} p\left( x \right), \\
 S_{q^*} \left( {\left. {\tilde X} \right|X} \right) =  - \sum\limits_x {\sum\limits_{\tilde x} {p\left( {x,\tilde x} \right) \ln _{q^*} p\left( {\left. {\tilde x} \right|x} \right)} } , \\
 S_{q^*} \left( {X,\tilde X} \right) =  - \sum\limits_x {\sum\limits_{\tilde x} {p\left( {x,\tilde x} \right) \ln _{q^*} p\left( {x,\tilde x} \right)} }  \\
=S_{q^*}(X)+S_{q^*}(\tilde X|X)=S_{q^*}(\tilde X)+ S_{q^*}(X|\tilde X),\\
and,\\
D_{K-L}^{q^*} \left[ {p\left( X \right)\left\| {r(X)} \right.}
\right] =  \sum\limits_x {p\left( x \right)}  \ln _{q^*}
\frac{{p(x)}}{{r(x)}},
\end{array}
\end{equation}
respectively. The dual Tsallis entropy has already been studied in
a maximum (Tsallis) entropy setting (for example, see Ref. [26]).
Note that the dual Tsallis entropy acquires a form identical to
the B-G-S entropies, with $ \ln_{q^*}(\bullet) $ replacing $
\log(\bullet) $. It is important to note that the $ q^* =2-q $
duality has been studied within the Sharma-Taneja-Mittal
framework by Kanniadakis, \textit{et. al.} [30]. The dual Tsallis entropy has been demonstrated to support a parametrically extended information theory, as is defined in Theorem 2 below. \\

\textbf{Theorem 1}~[2]:  \textit{Let $ X_1 ,X_2 ,X_3 ,...,X_n  $ be
random variables obeying the probability distribution $ p\left( {x_1
,x_2 ,x_3 ,...,x_n } \right) $, then  we have the chain rule
\begin{equation}
\begin{array}{l}
S_{q^*} \left( {X_1 ,X_2 ,X_3 ,...,X_n } \right) = \sum\limits_{i =
1}^n {S_{q^*} } \left( {\left. {X_i } \right|X_{i - 1} ,...,X_1 }
 \right). \\
\end{array}
\end{equation}}

\subsection{Generalized mutual informations}

Given a random variable $X$ in $\mathcal {X}$ where instances of $X$
are ${x_1,...,x_{|\mathcal{X}|}}$, for $ 0 < q < 1 $, the
generalized mutual information is defined in terms of the
generalized K-Ld [2]
\begin{equation}
I_{0<q<1} \left( {X;\tilde X} \right)  =  - \sum\limits_{x,\tilde x}
{p\left( {x,\tilde x} \right)\ln _q } \left( {\frac{{p\left( x
\right)p\left( {\tilde x} \right)}}{{p\left( {x,\tilde x} \right)}}}
\right).
\end{equation}

For nonadditivity parameters in the range $q>1$, the generalized
mutual information is [2,27]
\begin{equation}
\begin{array}{l}
 I_q \left( {X;\tilde X} \right) = S_q \left( X \right) - S_q \left( {\left. X \right|\tilde X} \right) = S_q \left( {\tilde X} \right) - S_q \left( {\left. {\tilde X} \right|X} \right) \\
  = S_q \left( X \right) + S_q \left( {\tilde X} \right) - S_q \left( {X,\tilde X} \right)=I_q(\tilde X;X); q > 1\\
 \end{array}
\end{equation}

For (18) to hold true, the inequalities (\textit{sub-additivities})
\begin{equation}
S_q \left( {\left. X \right|\tilde X} \right) \le S_q \left( X
\right),and,S_q \left( {\left. {\tilde X} \right|X} \right) \le S_q
\left( {\tilde X} \right),
\end{equation}
have to hold true.  This is not guaranteed for nonadditivity
parameters in the range $0<q<1$ [2,27].

As stated in Refs. [2] and [27], the generalized mutual
information is separately defined within  two separate $q-$ranges
 $ 0 < q < 1 $ and $ q > 1 $. They have different uses. For $ 0 <
q < 1 $, the generalized mutual information, as defined by (17),
provides a means of extrapolating the Csisz\'{a}r-Tusn\'{a}dy theory
[31] to the nonextensive domain for two convex sets of probability
distributions [2].  This has important implications in communication
theory and allied disciplines [2, 5].

For $ q > 1 $, the generalized mutual information as defined by
(18) possesses a number of important properties such as the
\textit{generalized data processing inequality} and the
\textit{generalized Fano inequality} [27].  This allows one to
define  Lagrangians and cost functions for processes defined by a
Markov chain relation.

\textbf{Theorem 2} [2] The generalized mutual information for
nonadditivity parameters in the range $0<q<1$ and $q>1$ are
related via the additive duality

\begin{equation}
\begin{array}{l}
I_{q^*}(X;\tilde X) =  - \sum\limits_x {\sum\limits_{\tilde x} {p\left( {x,\tilde x} \right)\ln _{q^ *  } } } \left( {\frac{{p\left( x \right)p\left( {\tilde x} \right)}}{{p\left( {x,\tilde x} \right)}}} \right) \\
\mathop   = \limits^{\left( q^* \rightarrow q \right)}S_q(X)+S_q(\tilde X)-S_q(X,\tilde X)=I_q(X;\tilde X);0<q^*<1,and,q>1. \\
\end{array}
\end{equation}

\subsection{Bregman divergences and scaled Bregman divergences}

This sub-section introduces the formal definition of Bregman
divergences and some of their select properties.  The Bregman
divergence or Bregman distance is similar to a metric, but does not
in general satisfy the triangle inequality nor symmetry. Bregman
divergences do however obey the Pythagorean theorem (for example,
see Appendix A in [12]). There are two ways in which Bregman
divergences are important. Firstly, they generalize squared
Euclidean distances to a class of distances that all share similar
properties. Secondly, they bear a strong connection to exponential
families of distributions. There is a bijection between regular
exponential families and regular Bregman divergences. Bregman
divergences are named after L. M. Bregman [10], who introduced the
concept in 1967. More recently researchers in geometric algorithms
have shown that many important algorithms can be generalized from
Euclidean metrics to distances defined by Bregman divergence. This
sub-section introduces the formal definition of Bregman divergences
and some of their properties.

\textbf{Definition 1} (Bregman divergences)[10, 32]: Let $ \phi $ be
a real valued strictly convex function defined on the convex set $
\mathcal{S} \subseteq dom(\phi) $, the domain of $ \phi $ such that
$ \phi $ is differentiable on $ ri(\mathcal{S}) $, the relative
interior of $ \mathcal{S} $. The Bregman divergence $ B_\phi
:\mathcal{S} \times {\mathop{\rm ri}} \left( \mathcal{S} \right)
\mapsto [0,\infty) $ is defined as: $ B_\phi \left( {z_1 ,z_2 }
\right) = \phi \left( {z_1 } \right) - \phi \left( {z_2 } \right) -
\left\langle {z_1  - z_2 ,\nabla \phi \left( {z_2 } \right)}
\right\rangle $, where: $\nabla \phi \left( {z_2 } \right) $ is the
gradient of $ \phi $ evaluated at $
z_2 $. \footnote{Note that $\left\langle\bullet,\bullet\right\rangle$ denotes the inner product.  Calligraphic fonts denote sets.}\\

\textbf{Definition 2} (Notations)[25]: $\mathcal{M}$ denotes the
space of all finite measures on a measurable space
$(\mathcal{X},\mathcal{A})$ and $\mathcal{P}\subset \mathcal{M}$ the
subspace of all probability measures. Unless otherwise explicitly
stated \textit{P},\textit{R},\textit{M} are mutually
measure-theoretically equivalent measures on
$(\mathcal{X},\mathcal{A})$ dominated by a $\sigma$-finite measure
$\lambda$ on $(\mathcal{X},\mathcal{A})$. Then the densities
\begin{equation}
p = \frac{{dP}}{{d\lambda }},r = \frac{{dR}}{{d\lambda }},and,m =
\frac{{dM}}{{d\lambda }},
\end{equation}
have a common support which will be identified with $\mathcal{X}$.
Unless stated otherwise, it is assumed that $P,R\in \mathcal{P},
M\in \mathcal {M}$ and that $\phi:(0,\infty)\mapsto \mathcal{R}$ is
a continuous and convex function.

\textbf{Definition 3} (Scaled Bregman Divergences) [25] \textit{The
Bregman divergence }of probability measures \textit{P}, \textit{R}
\textit{scaled} by an arbitrary measure M on
$(\mathcal{X},\mathcal{A})$ measure-theoretically equivalent with
\textit{P}, \textit{R} is defined by
\begin{equation}
\begin{array}{l}
 B_\phi  \left( {P,R\left| M \right.} \right) = \int_\mathcal{X} {\left[ {\phi \left( {\frac{p}{m}} \right) - \phi \left( {\frac{r}{m}} \right) - \left( {\frac{p}{m} - \frac{r}{m}} \right)\nabla \phi \left( {\frac{r}{m}} \right)} \right]} dM \\
  = \int_\mathcal{X} {\left[ {m\phi \left( {\frac{p}{m}} \right) - m\phi \left( {\frac{r}{m}} \right) - \left( {p - r} \right)\nabla \phi \left( {\frac{r}{m}} \right)} \right]} d\lambda . \\
 \end{array}
\end{equation}
The convex $\phi$ may be interpreted as the generating function of
the divergence. In a discrete setting, a scaled Bregman divergence
is defined as [25]

\begin{equation}
B_\phi  \left( {p,r\left| m \right.} \right) = \sum\limits_{i = 1}^d
{\left[ {\phi \left( {\frac{{p_i }}{{m_i }}} \right) - \phi \left(
{\frac{{r_i }}{{m_i }}} \right) - \left( {\frac{{p_i }}{{m_i }} -
\frac{{r_i }}{{m_i }}} \right)\nabla \phi \left( {\frac{{r_i }}{{m_i
}}} \right)} \right]m_i .}
\end{equation}

\section{Maximum dual Tsallis entropy models}

The Tsallis entropy parameterized by $q$ is defined as [1]
\begin{equation}
S_q \left[ p \right] =  - \frac{{1 - \sum\limits_i {p_i^q }
}}{{\left( {1 - q} \right)}}.
\end{equation}
Setting $q^*=2-q$, the dual Tsallis entropy is expressed as [2, 26]
\begin{equation}
S_{q^ *  } \left[ p \right] =  - \frac{{1 - \sum\limits_i {p^{2-q^
* } } }}{{\left( {q^ * -1 } \right)}}.
\end{equation}
The $q^*$-deformed Lagrangian (normal averages used)  to be
extremized reads
\begin{equation}
\Phi_* \left[ {p,\alpha ,\beta } \right] = S_{q^*} \left[ p \right]
- \alpha \left( {\sum\limits_i {p_{i} }  - 1} \right) - \beta \left(
{\sum\limits_i {p_{i} E}  - U} \right),
\end{equation}
yielding the canonical probability that maximizes the dual Tsallis
entropy as
\begin{equation}
\begin{array}{l}
 p_{i}  = \frac{{\left[ {1 - \frac{{\left( {1-q^ * } \right)}}{{\aleph _{q^ *  } }}\tilde \beta ^ *  \left( {E - U} \right)} \right]^{\frac{1}{{1-q^ *  }}} }}{{\aleph _{q^ *  }^{\frac{1}{{q^ *-1  }}} }}, \\
 \aleph _{q^ *  }  = \sum\limits_i {p_{i}^{2-q^ *  } }  = \tilde Z\left( {\tilde \beta ^ *  } \right)^{q^ *-1  } ,and,\tilde \beta ^ *   = \frac{\beta }{{2-q^ *  }} \\
 \end{array}
\end{equation}
Note that the methodology developed in [33] is employed in the
maximum Tsallis analysis using constraints defined by normal
averages. Here, $\tilde Z(\tilde\beta^*)$ is the canonical partition
function. The Appendix in this paper provides the detailed
derivation of (27).
 The dual Tsallis entropy is defined as
\begin{equation}
S_{q^*} \left[ p \right] = \frac{{\aleph _{q^*}  - 1}}{{\left( {
q^*-1} \right)}}.
\end{equation}
Here, $\tilde\beta^*=\beta/(2-q^*)$ is referred to as the "dual
scaled inverse thermodynamic temperature".
\section{Correspondence between the generalized Kullback-Leibler
divergences}

The generalized free energy (GFE) for normal averages expectations
is defined as [21]
\begin{equation}
F_q  = U - \frac{1}{{\tilde \beta }}S_q \left[ p \right].
\end{equation}
Note that the expression for the GFE (29) has recently been the
object of much research and debate.  The effective inverse
temperature $ \tilde \beta $ is the energy Lagrange multiplier
scaled with respect to $q$.  The energy Lagrange multiplier
generally relates to the thermodynamic temperature $ T $ as: $ \beta
= \frac{1}{{k_BT }} $, where $k_B$ is the Boltzmann constant
(sometimes set to unity for the sake of convenience) {\it only in
the limiting case} $q\rightarrow 1$. Prominent attempts to clarify
this issue are those by Abe \textit{et. al.} [34], Abe [35], amongst
others. Similarly, the $q^*$-deformed (dual) GFE is defined as
\begin{equation}
F_{q^ *  }  = U - \frac{1}{{\tilde \beta ^ *  }}S_{q^ *  } \left[ p
\right].
\end{equation}

At this stage, setting the reference probability $ r = \tilde p$ in
(9), and associating the quantities $ \tilde S_q ,\tilde
U,and,\tilde \aleph _q$ with the maximum Tsallis entropy canonical
distribution $ \tilde p_i$, yields
\begin{equation}
D_{q^ *  }^B \left[ {\left. p \right\|\tilde p} \right] = \frac{1}{{\left( {1-q^*} \right)}}\sum\limits_i {p_i } \left( {p_i^{1-q^*}  - \tilde p_i^{1-q^*} } \right) - \sum\limits_i {\left( {p_i  - \tilde p_i } \right)\tilde p_i^{1-q^*} }.  \\
\end{equation}

Substituting (27) into (31) one gets
\begin{equation}
\begin{array}{l}
 D_{q^ *  }^B \left[ {p\left\| {\tilde p} \right.} \right] = \frac{1}{{\left( {1 - q^ *  } \right)}}\sum\limits_i {p_i } \left( {\aleph _{q^ *  }  - \left( {1 - q^ *  } \right)\tilde \beta ^ *  U - \tilde \aleph _{q^ *  }  + \left( {1 - q^ *  } \right)\tilde \beta ^ *  \tilde U} \right) \\
  - \sum\limits_i {\left( {p_i  - \tilde p_i } \right)\left( {\tilde \aleph _{q^ *  }  + \left( {1 - q^ *  } \right)\tilde \beta ^ *  \left( {E - U} \right)} \right)}.  \\
 \end{array}
\end{equation}

With the aid of (28) and (30) and the normalization property, (32)
 leads now to

\begin{equation}
D_{q^ *  }^B \left[ {p_i \left\| {\tilde p_i } \right.} \right] =
\tilde \beta ^ *  \left[ { F_{q^ *  }  - \tilde F_{q^ *  } }
\right],
\end{equation}
and the dual generalized K-Ld defined in (15) becomes
\begin{equation}
\begin{array}{l}
 D_{K - L}^{q^ *  } \left[ {p_i \left\| {\tilde p_i } \right.} \right] =\sum\limits_i {p_i } \ln _{q^ *  } \left( {\frac{{p_i }}{{\tilde p_i }}} \right) \\
  = \frac{1}{{\left( {1 - q^ *  } \right)}}\sum\limits_i {p_i \left( {\frac{{p_i^{1 - q^ *  } }}{{\tilde p_i^{1 - q^ *  } }} - 1} \right)}  = \frac{1}{{\left( {1 - q^ *  } \right)}}\sum\limits_i {p_i \left( {\frac{{p_i^{1 - q^ *  }  - \tilde p_i^{1 - q^ *  } }}{{\tilde p_i^{1 - q^ *  } }}} \right)}  \\
  = \beta ^ *  \left[ {F_{q^ *  }  - \tilde F_{q^ *  } } \right]\tilde
\Psi _{q^ *  } \sum\limits_i {p_i }  = \beta ^ *  \left[ {F_{q^ *  }
- \tilde F_{q^ *  } } \right]\tilde \Psi _{q^ *  } ;\tilde \Psi _{q^
*  }  = \sum\limits_i {\tilde p_i^{q^ *   - 1} }.\\
\end{array}
\end{equation}

From (33) and (34), the correspondence relation between the usual
generalized K-Ld, the dual generalized K-Ld, and the generalized
Bregman K-Ld is

\begin{equation}
\begin{array}{l}
 D_{K - L}^B \left[ {p_i \left\| {\tilde p_i } \right.} \right] = \frac{{D_{K - L}^{q^ *  } \left[ {p_i \left\| {\tilde p_i } \right.} \right]}}{{\tilde \Psi _{q^ *  } }}\mathop  = \limits^{q^ *   \to q} \frac{{D_{K - L}^q \left[ {p_i \left\| {\tilde p_i } \right.} \right]}}{{\tilde \Psi _q }}; \\
 \tilde \Psi _q  = \sum\limits_i {\tilde p_i^{1 - q} },  \\
 \end{array}
\end{equation}
which is a compact result.

It is important to point out  that one application  of the
correspondence relation presented in this Section is that of
providing an alternate means to derive the dual generalized K-Ld
from the generalized Bregman K-Ld.  This may be accomplished by
invoking the linearity property of Bregman divergences (see
Appendix A of Ref. [12]).

The above mentioned linearity property states that the Bregman
divergence is a linear operator i.e., $\forall x \in \mathcal{S},y
\subset ri(S)$ (where $ri(\bullet)$ denotes the relative interior of
a set), $B_{c\phi}(x,y) = cB_{\phi}(x,y)$ (for c $>$ 0). From (35),
it is immediately evident that multiplying (31) by: $ c = \tilde
\Psi _{q^* }>0$ and invoking (12) readily yields the dual
generalized K-Ld. This relation between the dual generalized K-Ld
and Bregman divergences may however be viewed as one of convenience,
which although tenable, lacks the formal theoretical rigor of the
results presented in Section 5 below.

\section{Dual generalized K-Ld, scaled Bregman divergences, and the scaled Bregman information}

This Section serves a two-fold purpose: $(i)$ it is established that
the dual generalized K-Ld defined in (15) is a scaled Bregman
divergence, (ii) we introduce the concept of scaled Bregman
information as the expectation of a scaled Bregman divergence.

\subsection{Dual generalized K-Ld as a scaled Bregman divergence}

Let (i) $ t = \frac{z}{m}\,$ and (ii) the generating function of the
Bregman divergence be a convex function  $\phi(t)$,  with  $m$ the
scaling.  For a generating function $\phi(t)=t\, ln_{q^*}t$, the
discrete form of the scaled Bregman divergence (23) acquires the
form
\begin{equation}
\begin{array}{l}
 B_\phi  \left( {p,r\left| m \right.} \right) = \sum\limits_i {\left[ {\frac{{p_i }}{{m_i }}\ln _{q^ *  } \frac{{p_i }}{{m_i }} - \frac{{r_i }}{{m_i }}\ln _{q^ *  } \frac{{r_i }}{{m_i }} - \left( {\frac{{p_i }}{{m_i }} - \frac{{r_i }}{{m_i }}} \right)\nabla {\frac{{r_i }}{{m_i }}} \ln_{q^*} \left( {\frac{{r_i }}{{m_i }}} \right)} \right]} m_i . \\
  = \sum\limits_i {\left[ {p_i \ln _{q^ *  } \frac{{p_i }}{{m_i }} - p_i \ln _{q^ *  } \frac{{r_i }}{{m_i }} - \left( {p_i  - r_i } \right)\left( {\frac{{r_i }}{{m_i }}} \right)^{1 - q^ *  } } \right]}  \\
= \sum\limits_i {\left\{ {p_i m_i^{q^ *   - 1} \left[ {\ln _{q^ *  }
p_i  - \ln _{q^ *  } r_i } \right] - \left( {p_i  - r_i }
\right)\left( {\frac{{r_i }}{{m_i }}} \right)^{1 - q^ *  } }
\right\}}. \\
 \end{array}
\end{equation}

At this point, specifying $m_i=r_i$ in (36), and invoking (12) and
the normalization relation: $ \sum\limits_i {p_i  = \sum\limits_i
{r_i } = 1} $, the dual generalized K-Ld in (15) is recovered,
i.e.
\begin{equation}
B_\phi  \left( {p,r\left| m=r \right.} \right) = \sum\limits_i {p_i
\ln _{q^ *  } \left( {\frac{{p_i }}{{r_i }}} \right)}.
\end{equation}

This is a $q^*$-deformed f-divergence and is consistent with the
theory derived in Refs. [24] and [25], when extended to deformed
statistics.  The above result may also be employed in the case of
the dual generalized K-Ld between a conditional probability and a
marginal probability. Let $X$ and $Y$ be random variables in
$\mathcal{X}$ and $\mathcal{Y}$ respectively. Let the marginal
discrete probability measures be:$ \left\{ {p\left( {x_i } \right)}
\right\}_{i = 1}^n$ and $ \left\{ {p\left( {y_j } \right)}
\right\}_{j = 1}^m$, respectively.  In such circumstances, the dual
generalized K-Ld reads
\begin{equation}
D_{K - L}^{q^ *  } \left[ {p\left( {\left. Y \right|x_i }
\right)\left\| {p\left( {Y } \right)} \right.} \right] =
\sum\limits_j {p\left( {\left. y_j \right|x_i } \right)\ln _{q^ *  }
} \frac{{p\left( {\left. y_j \right|x_i } \right)}}{{p\left( {y_j }
\right)}},
\end{equation}
and is indeed a scaled Bregman divergence with the scaling: $ \,
p\left( {y_j} \right)$.

\subsection{Dual generalized K-Ld and the scaled Bregman information}

\textbf{Definition 4} [36]: For any Bregman divergence (or scaled
Bregman divergence) $ B_\phi :\mathcal{S} \times {\mathop{\rm int}}
\left( \mathcal{S} \right) \mapsto \Re ^ + $ and any random variable
$Z\sim w(z)$ (where $w(z)$ is the probability measure associated
with $Z$), $z\in \mathcal{Z}\subseteq \mathcal{S}$, \textbf{the
Bregman information} (or \textbf{scaled Bregman information})
\textit{which is a measure of the information in Z }is defined as
\begin{equation}
I_{\phi}(Z)=<B_{\phi}(Z,<Z>)>.
\end{equation}

For example, let $X$ be a random variable that takes values in
$\mathcal{X} = \left\{ {x_i } \right\}_{i = 1}^n $ following a
probability measure $p(\textbf{x})$. Let $\mu  = \langle X \rangle =
\sum\limits_i {p\left( {x_i } \right)} x_i$, and let $B_{\phi}$ be a
Bregman divergence (or scaled Bregman divergence). Then the Bregman
information (or scaled Bregman information) of $X$ is defined as
\begin{equation}
I_\phi  \left( X \right) = \sum\limits_i {p\left( {x_i } \right)}
B_\phi  \left( {x_i ,\mu } \right).
\end{equation}

Consider a random variable $Z_x$ which takes values in the set of
probability distributions: $\mathcal{Z}_x
 = \left\{ {p\left( {\left. Y \right|x_i } \right)} \right\}_{i = 1}^n
 $, following the marginal probability: $\left\{ {p\left( {x_i } \right)} \right\}_{i =
 1}^n$ defined over this set.  The expectation of $Z_x$ is
\begin{equation}
\mu  = \sum\limits_i {p\left( {x_i } \right)} p\left( {\left. Y
\right|x_i } \right) = \sum\limits_i {p\left( {x_i ,Y} \right)}  =
p\left( Y \right).
\end{equation}
Thus, from (38)-(41) the scaled Bregman information, which is the
dual generalized mutual information, may be defined as
\begin{equation}
\begin{array}{l}
 I_{q^ *  } \left( {X;Y} \right) = \sum\limits_i {p\left( {x_i } \right)} D_{K - L}^{q^ *  } \left[ {p\left( {\left. Y \right|x_i } \right)\left\| {p\left( Y \right)} \right.} \right] \\
  = \sum\limits_i {p\left( {x_i } \right)} \sum\limits_j {p\left( {\left. {y_j } \right|x_i } \right)\ln _{q^ *  } } \frac{{p\left( {\left. {y_j } \right|x_i } \right)}}{{p\left( {y_j } \right)}} \\
  = D_{K-L}^{q^*}[p(X,Y)||p(X)p(Y)] = I_\phi  \left( {Z_x } \right). \\
 \end{array}
\end{equation}
Similarly, the relation: $I_{q^*}(X;Y)=I_\phi(Z_y)$ also holds true,
when $Z_y$ is a random variable which takes values in the set of
probability distributions: $\mathcal{Z}_y
 = \left\{ {p\left( {\left. X \right|y_j } \right)} \right\}_{j =
 1}^m
 $, following the marginal probability: $\left\{ {p\left( {y_j } \right)} \right\}_{i =
 1}^m$ defined over this set.  In this case, the scaled Bregman divergence
 is:$D_{K - L}^{q^ *  } \left[ {p\left( {\left. X \right|y_j }
\right)\left\| {p\left( {X } \right)} \right.} \right] =
\sum\limits_i {p\left( {\left. x_i \right|y_j } \right)\ln _{q^ *  }
} \frac{{p\left( {\left. x_i \right|y_j } \right)}}{{p\left( {x_i }
\right)}}$, and the normal averages expectation is calculated with
respect to: $p(y_j)$.

For values: $q^*>1$, the scaled Bregman information acquires the
form
\begin{equation}
\begin{array}{l}
 I_\phi  \left( {Z_x } \right) = \sum\limits_{i,j} {p\left( {x_i ,y_j } \right)\ln _{q^ *  } \frac{{p\left( {\left. {y_j } \right|x_i } \right)}}{{p\left( {y_j } \right)}}}  \\
 \mathop  = \limits^{\left( a \right)} \sum\limits_{i,j} {p\left( {x_i ,y_j } \right)p\left( {y_j } \right)^{q^ *   - 1} \left( {\ln _{q^ *  } p\left( {\left. {y_j } \right|x_i } \right) - \ln _{q^ *  } p\left( {y_j } \right)} \right)}  \\
 \mathop  = \limits^{\left( b \right)}\sum\limits_{i,j} {p\left( {x_i ,y_j } \right)\sum\limits_i {p\left( {x_i } \right)^{q^ *   - 1} p\left( {\left. {y_j } \right|x_i } \right)^{^{q^ *   - 1} } } \left( {\ln _{q^ *  } p\left( {\left. {y_j } \right|x_i } \right) - \ln _{q^ *  } p\left( {y_j } \right)} \right)}  \\
 =\sum\limits_i {\sum\limits_j {p\left( {x_i } \right)p\left( {\left. {y_j } \right|x_i } \right)\sum\limits_i {p\left( {x_i } \right)^{q^ *   - 1} p\left( {\left. {y_j } \right|x_i } \right)^{^{q^ *   - 1} } } \left( {\ln _{q^ *  } p\left( {\left. {y_j } \right|x_i } \right) - \ln _{q^ *  } p\left( {y_j } \right)} \right)} }  \\
=\sum\limits_i {\sum\limits_j {p\left( {x_i ,y_j } \right)^{q^ *  } \left( {\ln _{q^ *  } p\left( {\left. {y_j } \right|x_i } \right) - \ln _{q^ *  } p\left( {y_j } \right)} \right)} }  \\
  =  - \sum\limits_j {p\left( {y_j } \right)^{q^ *  } \ln _{q^ *  } p\left( {y_j } \right) + \sum\limits_{i,j} {p\left( {x_i ,y_j } \right)^{q^ *  } \ln _{q^ *  } p\left( {\left. {y_j } \right|x_i } \right)} }  \\
  =  - \sum\limits_i {p\left( {x_i } \right)^{q^ *  } \ln _{q^ *  } p\left( {x_i } \right) + \sum\limits_{i,j} {p\left( {x_i ,y_j } \right)^{q^ *  } \ln _{q^ *  } p\left( {\left. {x_i } \right|y_j } \right)} }.  \\
 \end{array}
\end{equation}

In the derivation (43) $(a)$ denotes the use of (12) while $(b)$
denotes setting: $ p\left( {y_j } \right)^{q^ *   - 1}  =
\sum\limits_i {p\left( {x_i } \right)^{q^ *   - 1} } p\left( {\left.
{y_j } \right|x_i } \right)^{q^ *   - 1}$. Note that the last two
expressions in (43) are identical to (18) with the nonadditivity
parameter $q^*$ replacing  $q$  in (18). Defining
\begin{equation}
\begin{array}{l}
\tilde S_{q^*} \left( X \right) =  - \sum\limits_x {p\left( x
\right)} ^{q^*} \ln _{q^*} p\left( x \right), \\
\tilde S_{q^*} \left( {\left. {\tilde X} \right|X} \right) =  - \sum\limits_x {\sum\limits_{\tilde x} {p\left( {x,\tilde x} \right)^{q^*} \ln _{q^*} p\left( {\left. {\tilde x} \right|x} \right)} } , \\
\tilde S_{q^*} \left( {X,\tilde X} \right) =  - \sum\limits_x {\sum\limits_{\tilde x} {p\left( {x,\tilde x} \right)^{q^*} \ln _{q^*} p\left( {x,\tilde x} \right)} }  \\
=\tilde S_{q^*}(X)+\tilde S_{q^*}(\tilde X|X)=\tilde S_{q^*}(\tilde
X)+ \tilde S_{q^*}(X|\tilde X),\\,
\end{array}
\end{equation}
the scaled Bregman information (43) acquires the form
\begin{equation}
I_\phi  \left( {Z_x } \right)  = \tilde S_{q^
*  } \left( Y \right) - \tilde S_{q^ *  } \left( {Y\left| X \right.}
\right)= \tilde S_{q^ *  } \left( X \right) -\tilde S_{q^ *  }
\left( {X\left| Y \right.} \right)=I_\phi  \left( {Z_y } \right),
\end{equation}
where the inequalities: $\tilde S_{q^
* } \left( {\left. {Y} \right|X} \right) \le \tilde S_{q^ *  }
\left( {Y} \right)$, and, $\tilde S_{q^ *  } \left( {\left. X
\right|Y} \right) \le \tilde S_{q^ *  } \left( X \right)$ hold true.

Comparison of (13) and (44) readily reveals that the original
expressions for the Tsallis entropy and conditional Tsallis entropy,
and their equivalent forms derived from the dual generalized mutual
information (43), are invariant under interchange of the
nonadditivity parameters $q$ and $q^*=2-q$.  While this is indeed an
appealing observation, two points need to be noted: $(i)$ the
physics of the problem is defined by $q$ and not $q^*$, and, $(ii)$
Eqs. (13) and (44) correspond to two separate physical conditions.

The generalized mutual information (18) is expressed in terms of
(13) for $q>1$. This corresponds to probability distributions of
particular interest to Tsallis statistics, i.e. "long-tailed" and
power law distributions, amongst others.  On the other hand,
(43)-(45) correspond to $q^*> 1 \Rightarrow q < 1$. This regime is
not of great interest in generalized statistics. Thus, when modeling
problems in generalized statistics (for example, see Ref. [3]) whose
variational principle requires invoking the properties Bregman
divergences, use of Theorem 2 (Eq. (20)) is to be employed in order
to simultaneously achieve information-geometric and physical
consistency.

\section{Summary and Discussions}
Our present endeavors have enabled us to reach several findings
regarding the Tsallis environment.

\begin{itemize}

\item  The dual generalized K-Ld was shown to be a scaled Bregman
divergence.

\item With regards to expectation values computed using normal averages,
the dual generalized mutual information was demonstrated to be a
scaled Bregman information,

\item The correspondence linking  the
dual generalized K-Ld,  the generalized Bregman K-Ld (for
probability distributions which maximize the dual Tsallis entropy
when using normal-averages-constraints), and  the usual form of the
generalized K-Ld, has been established. Such a correspondence has
not been previously investigated in Tsallis statistics literature.
\end{itemize}

From the analyses in Sections 3-5, it becomes obvious from a
combined statistical physics plus information geometric
perspective that the dual generalized K-Ld should also be employed
as the measure of uncertainty when performing a minimum cross
entropy analysis (principle of minimum discrimination information)
[28, 29, 37] for constraints that employ normal averages.

\textit{A simpler justification stems from the fact that while in
the orthodox B-G-S theory the K-Ld is a Bregman divergence [12],
its Tsallis counterpart is not a Bregman divergence. Instead, as
established in this paper, the dual generalized K-Ld is a scaled
Bregman divergence}.
 Future work uses the results derived herein to analyze: $(i)$ the
generalized statistics rate distortion theory [2], $(ii)$ the
generalized statistics information bottleneck method [3] within the
context of scaled Bregman divergences and scaled Bregman
informations, and $(iii)$ deformed statistics extensions of the
minimum Bregman information principle and their applications in
machine learning [36].

\textbf{Acknowledgements}

RCV gratefully acknowledges support from \textit{RAND-MSR} contract
\textit{CSM-DI $ \ \& $ S-QIT-101155-03-2009}.

\newpage

\appendix
\numberwithin{equation}{section}
\renewcommand{\theequation}{A.\arabic{equation}}      

\section*{Appendix A:  Derivation of expression for the canonical  probability which maximizes the dual Tsallis entropy}

From (26), the maximum dual Tsallis entropy Lagrangian is
\begin{equation}
\Phi _{q^ *  } \left[ {p,\alpha ,\beta } \right] =  - \sum\limits_i
{p_i \ln _{q^ *  } p_i }  - \alpha \left( {\sum\limits_i {p_i  - 1}
} \right) - \beta \left( {\sum\limits_i {p_i E - U} } \right).
\end{equation}

Employing the stationarity condition: $ \frac{{\delta \Phi _{q^ *  }
\left[ {p,\alpha ,\beta } \right]}}{{\delta p_i }} = 0$ for each
$p_i$ and the normalization condition: $ {\sum\limits_i {p_i  = 1}
}$, yields
\begin{equation}
\begin{array}{l}
  - \frac{{\left( {2 - q^ *  } \right)}}{{\left( {1 - q^ *  } \right)}}p_i^{1 - q^ *  }  - \beta E - \alpha  = 0 \\
  \Rightarrow p_i  = \left[ {\frac{{\left( {1 - q^ *  } \right)}}{{\left( {q^ *   - 2} \right)}}\left( {\alpha  + \beta E} \right)} \right]^{\frac{1}{{1 - q^ *  }}}.  \\
 \end{array}
\end{equation}

Employing the Ferri-Martinez-Plastino methodology [33], the
normalization Lagrange multiplier $\alpha$ is obtained as follows.
Multiplying the first equation in (A.2) by $p_i$ and summing over
all indices $i$ yields
\begin{equation}
\alpha  = \left( {\frac{{\left( {q^ *   - 2} \right)}}{{\left( {1 -
q^ *  } \right)}}\sum\limits_i {p_i^{2 - q^ *  }  - \beta U} }
\right).
\end{equation}
Substituting (A.3) into (A.2) yields
\begin{equation}
\begin{array}{l}
 p_i  = \left[ {\left( {\sum\limits_i {p_i^{2 - q^ *  } }  - \frac{{\left( {1 - q^ *  } \right)}}{{\left( {2 - q^ *  } \right)}}\beta \left( {E - U} \right)} \right)} \right]^{\frac{1}{{1 - q^ *  }}}  \\
  = \aleph _{q^ *  }^{\frac{1}{{1 - q^ *  }}} \left[ {\left( {1 - \frac{{\left( {1 - q^ *  } \right)}}{{\left( {2 - q^ *  } \right)}}\frac{\beta }{{\aleph _{q^ *  } }}\left( {E - U} \right)} \right)} \right]^{\frac{1}{{1 - q^ *  }}} ;\aleph _{q^ *  }  = \sum\limits_i {p_i^{2 - q^ *  } }  \\
  \Rightarrow p_i  = \frac{{\left[ {\left( {1 - \frac{{\left( {1 - q^ *  } \right)}}{{\left( {2 - q^ *  } \right)}}\frac{\beta }{{\aleph _{q^ *  } }}\left( {E - U} \right)} \right)} \right]^{\frac{1}{{1 - q^ *  }}} }}{{\aleph _{q^ *  }^{\frac{1}{{q^ *   - 1}}} }}. \\
 \end{array}
\end{equation}
Thus (27) is derived.
\end{document}